# Internal wave mixing in warming Lake Grevelingen

**by Hans van Haren**

Royal Netherlands Institute for Sea Research (NIOZ) and Utrecht University, P.O. Box 59, 1790 AB Den Burg, the Netherlands.
e-mail: hans.van.haren@nioz.nl

ABSTRACT

Seasonal hypoxia or even anoxia can occur in some local deep basins of coastal waters. Such low summertime oxygen contents especially affect benthic life. The seasonal coastal hypoxia is commonly related to biological increased respiration and to physical limited vertical turbulent exchange that is associated with increased vertical stable density stratification. However, the same stratification can support internal waves that may break and locally generate turbulence. Here, we investigate the physics of internal wave motions in saltwater Lake Grevelingen (SW-Netherlands) during warming in mid-spring. Grevelingen is refreshed by weak tidal motions through an open sluice in its dam to the North Sea. The outer North Sea has a surface tidal range of about 3 m, but the lake surface tidal range is negligible (<0.1 m). To quantify vertical turbulent exchange, high-resolution temperature sensors were moored in conjunction with a current meter in 36 m water depth for three days. The site is known for anoxic conditions near the bottom in summer. While a 3-day, 30-m mean eddy diffusivity of $<[K_z]> = 4\pm2\times10^{-5}$ m$^2$ s$^{-1}$ is found, the overall mean turbulence dissipation rate ($\propto$turbulent flux) is $<[\varepsilon]> = 1.1\pm0.6\times10^{-7}$ m$^2$ s$^{-3}$. Turbulent mixing occurs episodically, via near-surface cooling during night and increased winds, via sparse shear-driven breaking of internal waves at the main pycnocline, and via sheared near-bottom currents. Shear-driven turbulence is not commonly found in fresh-water lakes. Just below the main pycnocline around mid-depth a layer of weak turbulence is observed, as in fresh-water lakes. The observed turbulent exchange is sufficient to warm the near-bottom waters over the course of summer, but insufficient to prevent the biology from over-consuming oxygen contents.

## 1. Introduction

Various natural and anthropogenic processes affect coastal areas more than the open ocean. The development of hypoxia in coastal bottom waters, usually in local deep basins, is an important effect of eutrophication (e.g., Diaz and Rosenberg, 2008; Howarth et al., 2011). Such near-bottom oxygen depletion occurs when the consumption during respiration exceeds the supply of oxygen-rich waters. The depletion typically develops seasonally as a result of enhanced biological activity in spring and summer. During that period of the year, the warming by solar insolation heats natural water bodies from above, which results in an increased vertical temperature difference when vertical turbulent mixing e.g. by wind and frictional flow over the sea floor, is reduced. Thus, increased vertical temperature difference is generally associated with the development of hypoxia (e.g., Hagens et al., 2015). The vertical temperature (in salt water also salinity) difference over a particular vertical distance determines the vertical density stratification. Enhanced turbulence near the surface, e.g. by wind, and near the sea floor will generate a stronger stratification in a relatively thin layer in the interior: the 'pycnocline', a result of interplay between potential and kinetic energy sources.

Further interplay or feed-back mechanism occurs as increased stratification can support more internal waves (IW), which may locally enhance turbulent diapycnal mixing and weaken stratification in the pycnocline when they break. IW are found everywhere in stably stratified natural water bodies like lakes and oceans, where they are generated via interaction of flow with sloping topography (e.g. Bell, 1977), via wave-wave interaction or after the passage of atmospheric disturbances including nighttime cooling (e.g., LeBlond and Mysak, 1978). The frequency range of freely propagating IW is between the inertial frequency f, the vertical component of the earth rotation, and the buoyancy frequency $N= (-g/\rho \cdot d\sigma_0/dz)^{-1/2}$. Here, g denotes the acceleration of gravity, $\rho$ a reference density and $\sigma_0 = \rho$ - 1000 + pressure correction the density anomaly referenced to the surface. Under particular conditions of wave interaction or deformation by vertical current shear, IW may break and episodically create vertical diapycnal turbulent exchange that is more than 100 times larger than molecular diffusion values (e.g., Gregg, 1989). In the open central North Sea, the summer stratification is thus found marginally stable to the point that IW-induced vertical mixing is sufficient to: 1. heat near-bottom waters through the course of summer, and, 2. supply sufficient nutrients from the near-bottom waters to the photic zone to sustain the late-summer phytoplankton bloom (van Haren et al., 1999).

In this paper, we investigate whether such open-sea-conditions of IW-breaking apply to coastal waters, in particular of Lake Grevelingen. The focus is on the quantification of the physics process of turbulent mixing.

Grevelingen is a relatively large saltwater lake in the southwest of the Netherlands. It has an area of 140 km$^2$ and mean and maximum depths of 5.4 and 48 m, respectively. The lake is a closed-off part of the Rhine-Meuse estuary, with at present a single sluice connection to the North Sea to its west. Since 1999, the sluice is open year round, except for some periods between September and December and during periods of extreme high waters (Wetsteyn, 2011). The salty North Sea water is denser than the fresher land run-off water and, after sinking to the bottom, yields a stable stratification in Grevelingen. This stratification is reinforced by solar insolation from above in spring, as warmer waters are less dense than relatively cool waters (for salinities larger than 23). The largest vertical density difference is usually found late-May/early-June (Wetsteyn, 2011; Hagens et al., 2015). Near-bottom temperature increases throughout summer.

For a lake, Grevelingen has some unusual characteristics, of which a weak tidal flow is the most important physics one. In 1965, the Grevelingen was closed-off by a dam from the tide-dominated North Sea that has a local crest-trough range of about 3 m. The single sluice in the dam still dampens virtually all of the lake surface tide leaving a range of about 0.1 m (Wetsteyn, 2011), but allows a weak tidal flow that can generate internal tidal motions of the vertical density stratification after interaction with the complex topography of Grevelingen lake-floor. Biochemically, its deeper parts become hypoxic, and occasionally even anoxic, every summer (peaking in August), although less so after relaxing the opening regime of the sluice since 1999 (Wetsteyn, 2011). The lowest oxygen contents are commonly observed well after the largest vertical density difference in May/June. The hypoxia results in particular microbiology bacterial behaviour and buffering capacity that are different from open-ocean and non-oxygen depleted coastal areas (Hagens et al., 2015). Recreationally, Grevelingen is important for surfing and sailing, and notably also for its diving with relatively clear waters in terms of visibility, for dutch standards.

Via the Brouwers-sluice the North Sea salt water enters Grevelingen in the southwest corner of the dam. (Plans are now made to include more sluices in the Brouwersdam in the near-future to increase the tidal effects). Behind the sluice the deepest point of Grevelingen is found. Near this point we deployed a mooring-cable of high-resolution temperature 'T'-sensors and a current meter for a few days during a measurement campaign end of May 2012. The mooring was designed to establish details of water motions in the interior, including IW-induced turbulence. Due to the large recreational pressure the mooring could not be left unattended and had to be recovered before the weekend.

## 2. Materials & methods

Observations were made using a single mooring consisting of a 30 m nylon-coated steel cable between a single top-buoy and an anchor-frame on 1 m high legs intended to lift instrumentation outside the very fluffy hypoxic mud layer. The net buoyancy of the mooring was 100 kg and the net anchor was about 150 kg. The top-heavy frame held a 1.2 MHz acoustic Doppler current profiler (ADCP), but it toppled over upon deployment by ship-pull and the ADCP did not function properly. Along the cable, 93 'NIOZ4' self-contained high-resolution T-sensors were taped every 0.33 m. The lowest sensor was, somewhat unintentionally, at 0.4 m above the bottom. The sensors sampled at a rate of 2 Hz, with a precision better than $5\times10^{-4}$ °C and a noise level of less than $1\times10^{-4}$ °C (van Haren, 2018). A small five vertical 0.5-m-bins upward looking ADCP, a Teledyne/RDI Doppler Volume Sampler current profiler, was clamped to the cable at about 15 m. It sampled at a rate of once per 3 s, but due to battery problems it gave data for the first 1.4 days of deployment only.

The mooring was in the water for almost 3 days from 29 May 10 UTC to 1 June 7 UTC 2012 at 51° 44.58′ N, 3° 50.42′ E, 36 m water depth. The mooring site is about 1 km East from the sluice, in a relatively deep channel running East-West (Fig. 1). A longer mooring period was not possible because of the extensive recreational activities over the weekend and the lack of guard-buoys. The first two days were characterized by above-average warm and weak wind conditions. The third day showed below-average cool weather, with some wind and a little rain. While no meteorological data are available from the mooring site, hourly data are available from Royal Netherlands Meteorological Institute KNMI-stations Wilhelminadorp (25 km to the SSE of the mooring) and Vlissingen (40 km to the SSW).

On 30 May, and previously on 24 May, profiles of shipborne Conductivity-Temperature-Depth (CTD) were obtained close to the mooring site, including profiles from additional sensors like on oxygen content. The CTD-profiles were used for calibration guidance of the moored T-sensors and to establish a local temperature-salinity, or more precisely, a temperature-density relationship. Only if this relationship is tight, moored T-sensor data can be used as tracer for density to quantify turbulence parameter values.

Every 4 hours all T-sensors were synchronized to a standard clock, so that their times were less than 0.02 s off. Five sensors showed calibration or electronic problems and their data were linearly interpolated. After calibration and correction for electronic drift effects of less than 1 mK/mo using the CTD-data for guidance, the T-sensor data were converted into 'Conservative' (~potential) Temperature data $\Theta$ (IOC, SCOR, IAPSO, 2010). This corrects for compressibility effects. With reference to $\sigma_\theta$ the CTD-data provided a stable relation $\delta\sigma_\theta = \alpha\delta\Theta$ with $\alpha = -0.33\pm0.01$ kg m$^{-3}$ °C$^{-1}$ denoting the temperature-density slope under local conditions. This tight linear relation allows T-sensor data to be used as tracer for density variations, noting that salinity contributes positively to density variations.

Vertical turbulent kinetic energy dissipation rate $\varepsilon$, which is proportional to turbulent diapycnal flux, and eddy diffusivity $K_z$ are estimated by calculating overturning scales. These scales are obtained after reordering every 2-Hz time-step the potential density (temperature) profile $\rho(z)$, which may contain inversions, into a stable monotonic profile $\rho(z_s)$ without inversions (Thorpe, 1977). Here $z_s$ denotes the nearest vertical position away from position $z$ for the density profile to become stable. After comparing raw and reordered profiles, displacements $d = \min(|z-z_s|)\cdot\mathrm{sgn}(z-z_s)$ are calculated necessary for generating the stable profile. Then,

$$\varepsilon = 0.64d^2N^3, \tag{1}$$

where N denotes the buoyancy frequency computed from the reordered profile. The constant follows from empirically relating the overturning scale with the largest energy containing Ozmidov-scale $L_O$, which results in root mean squared $<L_O/d>_{rms} = 0.8$ (Dillon, 1982), a mean coefficient value from many realizations. Using $K_z = \Gamma\varepsilon N^{-2}$ and a mean mixing efficiency coefficient for conversion of kinetic into potential energy of $\Gamma = 0.2$ for ocean, not necessarily laboratory, observations (Osborn, 1980; Oakey, 1982; Gregg et al., 2018), we find,

$$K_z = 0.128d^2N. \tag{2}$$

According to Thorpe (1977), results from (1) and (2) are only useful after averaging over the size of an overturn. In the following, 'sufficient' averaging is applied over at least vertical scales of the largest overturns and over at least buoyancy time scales to warrant a concise mixture of convective- and shear-induced turbulence, and to justify the use of the above mean coefficient values. Due to the small precision of the T-sensors, thresholds limit mean turbulence parameter values to $<\varepsilon>_{thres} = O(10^{-12})$ m$^2$s$^{-3}$ and to $<K_z>_{thres} = O(10^{-6})$ m$^2$s$^{-1}$ in weakly stratified waters (van Haren et al., 2015). Here and in the following, averaging over time is denoted by […], averaging over depth-range by $<…>$. Molecular diffusion has values of about $10^{-7}$ m$^2$ s$^{-1}$ in water, while weak open-ocean and lake-interior dissipation rates have values $O(10^{-10}-10^{-9})$ m$^2$ s$^{-3}$ (e.g., Gregg, 1989; Wüest and Lorke, 2003). In comparison with multiple shipborne shear- and temperature-variance microstructure profiling the present method yielded similar results to within a factor of two.

## 3. Results and discussion

### 3.1. Vertical profile overview

The CTD- and time-mean T-sensor profiles yield the main result of this paper in a glance (Fig. 2). End of May, the daytime warming is seen to increase the near-surface temperature by about 3°C in the 6 days between the CTD-profiles (Fig. 2a). As the near-bottom temperature does not vary much, the vertical stratification increases with time as well. The 3-day mean T-sensor data fit the 30 May CTD-profile rather well, including a moderate stratification in the near-surface layer down to about z = -12 m, a rather strongly stratified pycnocline between -16 < z < -12 m, and a weak but non-negligible stratification deeper down.

The temperature stratification is largely followed by the salinity stratification (Fig. 2b), with saltier waters in the deeper layer confirming the notion that both temperature and salinity gradients contribute positively to vertical density variations and of approximately the same value. (The temperature/salinity coefficient ratio is approximately 4.5). In contrast with temperature stratification increasing with time, salinity stratification decreases between the two CTD-profiles, the near-surface waters becoming relatively saltier and the near-bottom waters becoming relatively fresher in the 6-day period between the CTD-profiles. While the former may be due to net evaporation, the latter can only be the result of internal turbulent exchange in a closed basin. It is noted however that the CTD-profiles are single snapshots in time, not providing insight of variation over a tidal cycle, for example.

Like temperature and salinity, oxygen (Fig. 2c) shows a vertical gradient with oxygen-rich waters in the upper 10 m near the surface and partially depleted waters below. End of May no anoxic conditions were observed, but the saturation level is below 50% near the bottom and decreasing in the 6-day period.

The amount of internal turbulent exchange expressed in the 3-day time-mean turbulence parameter profiles (Fig. 2d) shows highest values in the near-surface layer and the pycnocline. Values of about $10^{-7}$ m$^2$ s$^{-3}$ for dissipation rate are relatively high, well exceeding those of the open-ocean, and comparable with values for internal wave breaking above steep deep-ocean topography (van Haren et al., 2015). Eddy diffusivity values are comparable with open-ocean values of a few times $10^{-5}$ m$^2$ s$^{-1}$ (Gregg, 1989), presumably because of the relatively strong stratification still yielding a relatively large turbulent flux. In the present data, near-bottom z < -25 m mean values are similar in diffusivity and an order of magnitude smaller in dissipation rate due to the weaker stratification, in comparison with those from the near-surface layer. Between -25 < z < -18 m a layer of relatively weak turbulence is observed, with values approaching open-ocean and/or molecular values. This suggests very limited exchange between the waters up to 10 m from the bottom and those down to 18 m below the surface, during this period.

*3.2. Temperature overview*

The 3-day time and 30.5-m vertical overview of Conservative Temperature shows a variation of the temperature profiles on various scales (Fig 3a). A tidal variation is visible around z = -15 m, at the bottom of the pycnocline. Its vertical variation is 2 to 3 m, much larger than the 0.1 m lake surface tidal range. In contrast, a few meters higher-up, e.g. around z = -10 m, tidal excursions are not in parallel with those at z = -15 m, if discernible at all. Clearly, the tidal motions are not driven by surface excursions 'barotropic tide' but internally 'baroclinic tide'. A daily warming is observed, but not very distinctly also because the upper sensor was still about 5 m from the surface. Superposed on the tidal and daily variations are needle thin short-scale excursions, which will become identified as packets of high-frequency internal waves in magnifications below.

A hint of these short-scale excursions is given in the form of the first derivative of the vertically averaged T-sensor data (Fig. 3b). The apparent spikes are fast-fluctuating temperature variations, mostly due to smooth waves of 200-400 s periods, some are due to irregular overturns. The spikes show relatively low activity during nighttime and most activity during daytime, with an extension into the night of 1 June (yearday 152).

To get a flavour of the details of stratification variations with time and depth, the overall rather smooth vertical temperature and salinity profiles of Fig. 2 are seen to be organized in much smaller layering (Fig. 3c). The layering varies in thickness, vertical position and with time. As far as can be established with sensors 0.33 m apart, layer thickness varies from 0.33 m up to about 5 m, in these data. The large stratification is indeed found at the depth of the main pycnocline, weakly varying with the tidal cycle, at the end of 31 May (around day 151.8) when rain and cooling create relatively weak stratification in the upper 10 m due to turbulence 'convection'. The deformation of the stratification is due to vertical current shear, predominantly due to straining of various IW-packets. IW-straining is especially also modifying the layering of the weaker stratification for z < -20 m.

The abundant IW-activity and straining results in episodic breaking and turbulent patches (Fig. 3d) that result in the weaker stratification parts of Fig. 3c. The un-averaged turbulence dissipation rate data demonstrate the size of the patches in time-depth, including the (lower part of) convection during the rain-cooling around day 151.8 and a relatively large near-bottom value in the beginning of the record. While a particular periodicity of the mixing patches is difficult to establish across the panel, some tidal periodicity is seen very near (z < -33 m) the bottom. This may be due to tidal friction and/or straining. Vertical and time mean turbulence values for Fig. 3 are $[<\varepsilon>] = 1.1\pm0.6\times10^{-7}$ m$^2$s$^{-3}$, $[<K_z>] = 4\pm2\times10^{-5}$ m$^2$s$^{-1}$, with $[<N>] = 3.0\pm0.4\times10^{-2}$ s$^{-1}$ providing a mean buoyancy period of 210 s, and the minimum small-, 0.33-m scale buoyancy period is 44 s.

### 3.3. Some environmental variation

Days 149 and 150 showed relatively warm air increasing the upper T-sensor data weakly during daytime (Fig. 4a). Bright solar insolation (Fig. 4b) was followed by a rather cloudy day, with less than half the previous global radiation and some rain around day 151.8. While every day wind speed (Fig. 4b, red) increased from negligible night-values to about 5 m s$^{-1}$, it increased to 11 m s$^{-1}$ on day 151.6. This, together with the lesser radiation and rain, resulted in the cooling due to turbulent convection down to z = -10 m between days 151.4 and 151.8 (Fig. 4c). The stronger winds were basically from the West, hence more or less directed along the channel.

The previously noted high-frequency IW-activity is visible in Fig. 4c in the various isotherms, mainly down to about z = -20 m, with some lower-frequency motions extending deeper. While this activity was related to daytime heating as it started about 2 h after sunrise, it is seen here that wind speed forces it to be more intense in amplitude and of longer duration on day 151. A tidal modulation of these high-frequency IW is not observed in this short data-set.

The along-channel current component u (Fig. 5c red) and stratification (Fig. 5d purple shows N) are dominated by the tidal variation at the depth of the pycnocline between about -15 < z < -12 m. However, the cross-channel current component v (Fig. 5c, blue), the turbulence dissipation rate (Fig. 5b) and the vertical current shear-magnitude $|S| = ((du/dz)^2+(dv/dz)^2)^{1/2}$ (Fig. 5d green) are dominated by non-tidal variations with periodicities of a few hours. These latter three parameters also not vary with diurnal solar insolation, and hence not with high-frequency IW appearance. The shear-magnitude $|\mathbf{S}|$ does not correspond with variations in N, which is different from open-ocean findings at ten times larger vertical scale resolution (Pinkel and Anderson, 1997). The $|\mathbf{S}|$-variation with time dominates variations in the gradient Richardson number $Ri = N^2/|S|^2$ (Fig. 5e shows the inverse 1/Ri). Below certain thresholds $Ri_c$, for linear stability $Ri_c = 0.25$ (Miles 1961; Howard, 1961) and for stability in three-dimensional flows $Ri_c = 1$ (Abarbanel et al., 1984), destabilizing shear overcomes stabilizing stratification to induce turbulent overturning. It is seen that a reasonable correspondence is observed between turbulence dissipation rate (Fig. 5b) and 1/Ri (Fig. 5e). This suggests that most diapycnal turbulent exchange is driven by IW-shear, but not all.

### 3.4. Typical magnifications

The three-day record shows distinct episodes in which turbulent overturning is driven by buoyant convection, when denser water is generated or moved over less dense water. These are found near the surface during nighttime, most out of reach of the T-sensor array, and on day 151 when cooler and moist air blew over Grevelingen. The convective overturning reached down to z = -10 m (Fig. 6a). The overturns have time-scales of several 100 s. They force the stratification into a relatively thin layer and generate interfacial waves. These waves are very irregular and seldom sinusoidal in shape.

Much more regular sinusoidal smooth waves come in packets of 'solitary' waves, for example on day 150 (Fig. 6b). Initiated as local mode-2 waves with upper isotherms propagating 180° out-of-phase with lower isotherms, the high-frequency IW of about 150 s periodicity near the local buoyancy period transfer within one and a half periods in local mode-1 waves (when all isotherms propagate in-phase). The former are known for the enclosed core and potential turbulence contribution (Lamb, 2002; Preusse, 2012), the latter are typical for very linear, sparsely breaking waves. They contribute little to the turbulence dissipation rate.

An example of relatively large shear-induced overturning at the lower side of the pycnocline if given in Fig. 6c. Two so-called Kelvin-Helmholtz instabilities are about 5 h apart and last about 1 h. This is much larger than the mean buoyancy period of 200 s, but compares with the local minimum stratification having buoyancy periods of up to 6000 s. Similar overturning is also observed in the weakly stratified near-bottom layer (Fig. 6d). The overall temperature range is about ten times less than in Fig. 6c. The isotherm deformations are mainly shear-induced in both cases.

## 4. Concluding discussion

The present observations provide an overview of various IW-propagation and -breaking inducing turbulent overturning in saltwater lake Grevelingen. While the along-channel flow pycnocline variations are tide-dominated, high-frequency IW-generation and turbulent overturning are not predominantly varying with the tide. Other mechanisms prevail. IW near the buoyancy frequency vary with the local stratification, which partially varies with diurnal periodicity, and are found most intense during daytime. Atmospheric influences of wind-stress and cooling affect the stratification intensity of the pycnocline, by modifying the length-scale of largest density variation, and add to the intensity of the high-frequency IW. As the T-data were from a single mooring, IW-propagation could not be established. From previous observations and modelling, high-frequency IW are suggested to be generated by cross-channel flow generating local mode-2 followed by propagation into a train of mode-1 waves after the turn of the tide (Xie et al., 2017). As no tidal periodicity is observed, the generator is more likely atmosphere-related here.

In an enclosed fresh-water lake, IW-packets are generally found to propagate along the channel while dissipating their energy along the sides and at the far terminal end (Preusse, 2012). Diffusivity values are ten times higher than the near-molecular values found in the interior as established from tracer release experiments (Goudsmit et al., 1997). In Grevelingen, indeed little turbulence generation is observed in the high-frequency IW that mainly propagate as linear waves. Nevertheless, episodic large turbulence mixing is found across the pynocline between -15 < z < -10 m, and in the 10 m nearest to the bottom. Turbulence dissipation rate values are commensurate with those typical for lakes very close to the surface and bottom (Wüest and Lorke, 2003).

In Grevelingen, IW-breaking occurs predominantly shear-driven, and is associated with 1 to 5 h periodic motions, in the middle of the IW-frequency band [f, N]. This contrasts with common, generally fresh-water lakes in which turbulence is more convective- than shear-driven (Etemad-Shahidi and Imberger, 2001). A notable difference between salt-water lake Grevelingen and fresh-water lakes is a lack of tidally varying external forces in the latter. Although the precise mechanism behind the shear-driven IW-breaking is not known, the observations suggest nonlinear generation of higher (tidal) harmonics from below, and possibly atmospherically induced high-frequency IW from above. This requires further investigation.

In some aspects, Grevelingen behaves similar to the open central North Sea. Both stratify seasonally by the varying solar insolation, both show a monotonic increase in near-bottom temperature throughout the summer-stratified period (van Haren et al., 1999; Hagens et al., 2015) and both show similar turbulent diffusivity values across the pycnocline of $K_z = $ 3-5×10$^-$

$^5$ m$^2$ s$^{-1}$, several 100 times larger than molecular diffusion. Reduced turbulence caused the spring diatom bloom to sink out of the photic zone in the central North Sea (van Haren et al., 1998). It is not known whether that equally applies to Grevelingen, potentially causing a phytoplankton-rich near-bottom layer. However, diatoms are known to occur in Grevelingen (Rijstenbil, 1987), often in combination with *Phaeocystis* (Peperzak et al., 1998) that is considered harmful and potentially deteriorates near-bottom oxygen conditions (Wetsteyn, 2011).

In Grevelingen, the layer between -25 < z < -20 m of weak turbulence, with exchange rates down to molecular levels despite the ubiquitous internal waves, may be crucial for the set-up of hypoxic near-bottom waters later in summer. It is to be investigated whether this regime will change when multiple sluices are installed in the Brouwersdam in the future. The more permanent opening of the present sluice since 1999 allowed more salty water in, thereby increasing the vertical density difference but also allowing tidal and other IW-motions. It may equally be investigated what surface temperature changes have on (the length-scale of) stratification, the potential feed-back mechanism of IW-generation and -breaking, and on hypoxia levels.

**Acknowledgments**

I thank the captain and the crew of the R/V Luctor for their kind assistance during the overboard operations. F. Meysman organized the cruises and provided the CTD-data. I thank M. Laan for his enthusiastic development of NIOZ-temperature sensors. NIOZ T-sensors have been funded in part by the Netherlands Organization for the Advancement of Science, N.W.O.

**Declarations of interest**

None.

**Role of funding source**

The funding source had no involvement in study design, collection, analysis and interpretation of data as well as in writing the report or the decision to submit the article.

**References**

Abarbanel, H.D.I., Holm, D.D., Marsden, J.E., Ratiu, T., 1984. Richardson number criterion for the nonlinear stability of three-dimensional stratified flow. Physical Review Letters 52, 2352-2355.

Bell, T.H., 1975. Topographically generated internal waves in the open ocean. Journal of Geophysical Research 80, 320-327.

Diaz, R.J., Rosenberg, R., 2008. Spreading dead zones and consequences for marine ecosystems. Science 321, 926-929.

Dillon, T.M., 1982. Vertical overturns: a comparison of Thorpe and Ozmidov length scales. Journal of Geophysical Research 87, 9601-9613.

Etemad-Shahidi, A., Imberger, J., 2001. Anatomy of turbulence in thermally stratified lakes. Limnology and Oceanography 46, 1158-1170.

Goudsmit, G.-H., Peeters, F., Gloor, M., Wüest, A., 1997. Boundary versus internal diapycnal mixing in stratified natural waters. Journal of Geophysical Research 102, 27,903-27,914.

Gregg, M.C., 1989. Scaling turbulent dissipation in the thermocline. Journal of Geophysical Research 94, 9686-9698.

Gregg, M.C., D'Asaro, E.A., Riley, J.J., Kunze, E., 2018. Mixing efficiency in the ocean. Annual Review of Marine Science 10, 443-473.

Hagens, M., et al., 2015. Biogeochemical processes and buffering capacity concurrently affect acidification in a seasonally hypoxic coastal marine basin. Biogeosciences 12, 1561-1583.

Howard, L.N., 1961. Note on a paper of John W. Miles. Journal of Fluid Mechanics 10, 509-512.

Howarth, R., et al., 2011. Coupled biogeochemical cycles: eutrophication and hypoxia in temperate estuaries and coastal marine ecosystems. Frontiers in Ecology and the Environment 9, 18-26.

IOC, SCOR, IAPSO, 2010. The international thermodynamic equation of seawater – 2010: Calculation and use of thermodynamic properties. Intergovernmental Oceanographic Commission, Manuals and Guides No. 56, UNESCO, Paris, France, 196 pp.

Lamb, K., 2002. A numerical investigation of solitary internal waves with trapped cores formed via shoaling. Journal of Fluid Mechanics 451, 109-144.

LeBlond, P.H., Mysak, L.A., 1978. Waves in the Ocean. Elsevier, 602 pp.

Miles, J.W., 1961. On the stability of heterogeneous shear flows. Journal of Fluid Mechanics 10, 496-508.

Oakey, N.S., 1982. Determination of the rate of dissipation of turbulent energy from simultaneous temperature and velocity shear microstructure measurements. Journal of Physical Oceanography 12, 256-271.

Osborn, T.R., 1980. Estimates of the local rate of vertical diffusion from dissipation measurements. Journal of Physical Oceanography 10, 83-89.

Peperzak, L., Colijn, F., Gieskes, W.W.C., Peeters, J.C.H., 1998. Development of the diatom-*Phaeocystis* spring bloom in the Dutch coastal zone of the North Sea: the silicon depletion versus the daily irradiance threshold hypothesis. Journal of Plankton Research 20, 517-537.

Pinkel, R., Anderson, S., 1997. Shear, Strain, and Richardson Number Variations in the Thermocline. Part I: Statistical Description. Journal of Physical Oceanography 27, 264-281.

Preusse, M., 2012. Properties of internal solitary waves in deep temperate lakes. PhD-thesis, University of Konstanz, D, 114 pp.

Rijstenbil, J.W., 1987. Phytoplankton composition of stagnant and tidal ecosystems in relation to salinity, nutrients, light and turbulence. Netherlands Journal of Sea Research 21, 113-123.

Thorpe, S.A., 1977. Turbulence and mixing in a Scottish loch. Philosophical Transactions of the Royal Society of London A 286, 125-181.

van Haren, H., 2018. Philosophy and application of high-resolution temperature sensors for stratified waters. Sensors 18, 3184, doi:10.3390/s18103184.

van Haren, H., Mills, D.K., Wetsteyn, L.P.M.J., 1998. Detailed observations of the phytoplankton spring bloom in the stratifying central North Sea. Journal of Marine Research 56, 655-680.

van Haren, H., Maas, L., Zimmerman, J.T.F., Ridderinkhof, H., Malschaert, H., 1999. Strong inertial currents and marginal internal wave stability in the central North Sea. Geophysical Research Letters 26, 2993-2996.

van Haren, H., Cimatoribus, A.A., Gostiaux, L., 2015. Where large deep-ocean waves break. Geophysical Research Letters 42, 2351-2357, doi:10.1002/2015GL063329.

Wetsteyn, L.P.M.J., 2011. Grevelingenmeer: meer kwetsbaar? Een beschrijving van de ecologische ontwikkelingen voor de periode 1999 t/m 2008-2010 in vergelijking met de periode 1990 t/m 1998. RWS Waterdienst, Lelystad, the Netherlands, 163 pp (in Dutch).

Wüest, A., Lorke, A., 2003. Small-scale hydrodynamics in lakes. Annual Review of Fluid Mechanics 35, 373-412.

Xie, X., Li, M., Scully, M., Boicourt, W.C., 2017. Generation of internal solitary waves by lateral circulation in a stratified estuary. Journal of Physical Oceanography 47, 1789-1797.

**Fig. 1.** Bathymetric map of Lake Grevelingen, a former estuarine sea-arm in the southwest of the Netherlands. The mooring- and CTD-site is indicated by the red triangle.

**Fig. 2**. Vertical profile overview of water characteristics end of May 2012. (a) Conservative (~potential) Temperature, comparing three-day time mean moored T-sensor data (thick solid graph) with single shipborne CTD-profiles obtained on 24 May (dotted) and 30 May (dashed). (b) Absolute Salinity for the two CTD-profiles. (c) Oxygen saturation for the two CTD-profiles. (d) Logarithm of three-day time mean turbulence dissipation rate (dots; top-scale) and eddy diffusivity (solid graph; bottom-scale) estimated from moored T-sensor data.

**Fig. 3**. Moored T-sensor data overview of three-day time or time-depth series. (a) Conservative Temperature. The horizontal axis is at the local lake floor. Rectangles refer to detailed plots in Fig. 6a-d. (b) Absolute first derivative of vertical mean values of the data in a. (c) Logarithm of the buoyancy frequency determined from the data in a. after reordering to statically stable profiles. (d) Logarithm of turbulence dissipation rate computed from the data in a.

**Fig. 4**. Comparison of nearby weather station information with moored T-sensor data. (a) Temperature measured in water by the upper T-sensor at 5 m below the surface (solid graph) and in air at Wilhelminadorp 'WD' station (dashed). (b) Solar global radiation at WD (dashed black, left-scale) and wind speed (red, right-scale) at WD (dashed) and Vlissingen station (solid). The vertical bars indicate times of local sunrise (solid) and sunset (dashed). (c) As Fig. 3a, but with color-scale focusing on the lower half of the water column and including contours of constant temperature. Solid contours are drawn every 0.05°C, white (distinguishing the pycnocline) and purple (upper layer) contours every 0.5°C.

**Fig. 5**. Time series of the first 1.4 days of data when the current profiler at z = -15 m was working. Current meter data are smoothed to 60-s-data for noise removal using a double elliptic filter.

(a) Conservative Temperature from T-sensor at z = -14.5 m. (b) Logarithm of turbulence dissipation rate vertically averaged over -16.2 < z < -12.9 m. (c) Along-channel east-west horizontal current component in red and cross-channel component in blue at z -14.5 m. (d) Buoyancy frequency N from vertically reordered T-sensor data and shear-magnitude |**S**| between -14 < z < -13.5 m. (e) Logarithm of 1/Gradient Richardson number from the data in d., see text. Critical values are indicated by horizontal lines. In this plot unstable portions are above the critical lines.

**Fig. 6**. Detail examples of T-sensor data, see Fig. 3 for time-depth ranges. Every panel has different Conservative Temperature (colour) range. Ten black contours are evenly distributed over the temperature range per panel. (a) Near-surface turbulence due to night-time cooling. The time, vertical range of the panel is [1.2 h, 7 m]. (b) High-frequency interfacial wave packet with frequencies near the local buoyancy frequency, [2300 s, 14 m]. (c) Episodic interfacial overturning, [9 h, 6.5 m]. (d) Occasional near-bottom overturning, [6 h, 13 m].

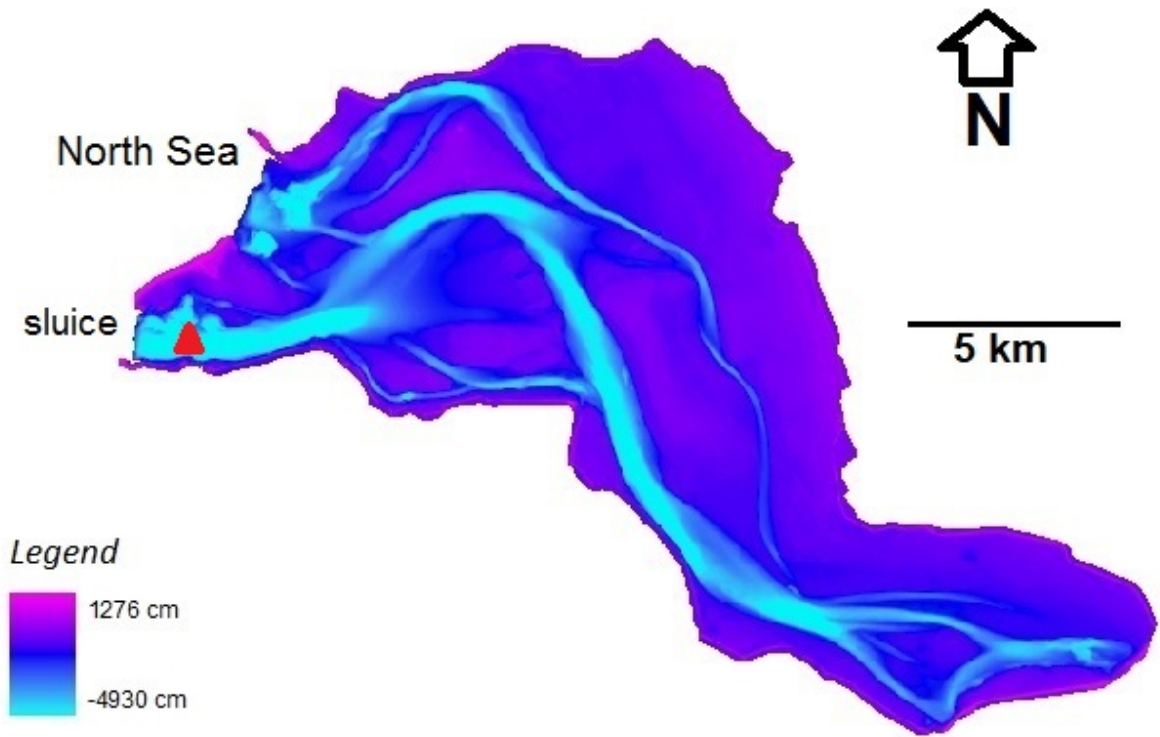

**Fig. 1.** Bathymetric map of Lake Grevelingen, a former estuarine sea-arm in the southwest of the Netherlands. The mooring- and CTD-site is indicated by the red triangle.

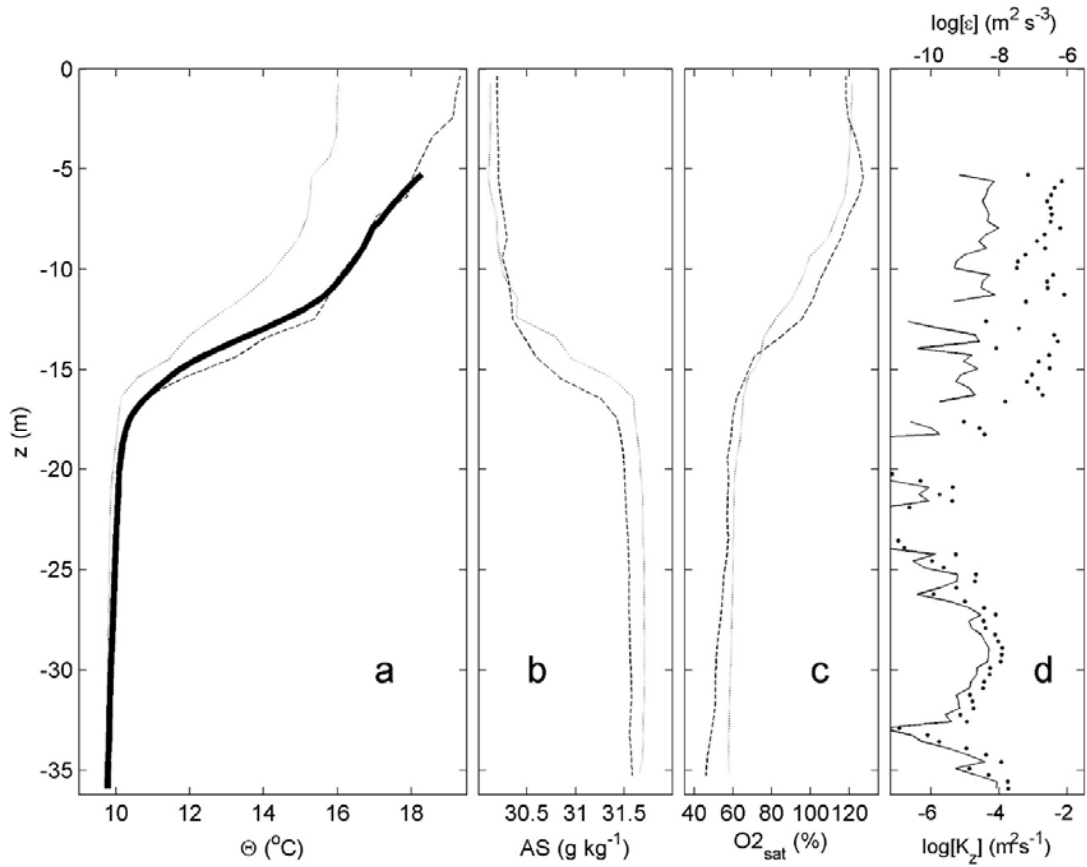

**Fig. 2**. Vertical profile overview of water characteristics end of May 2012. (a) Conservative (~potential) Temperature, comparing three-day time mean moored T-sensor data (thick solid graph) with single shipborne CTD-profiles obtained on 24 May (dotted) and 30 May (dashed). (b) Absolute Salinity for the two CTD-profiles. (c) Oxygen saturation for the two CTD-profiles. (d) Logarithm of three-day time mean turbulence dissipation rate (dots; top-scale) and eddy diffusivity (solid graph; bottom-scale) estimated from moored T-sensor data.

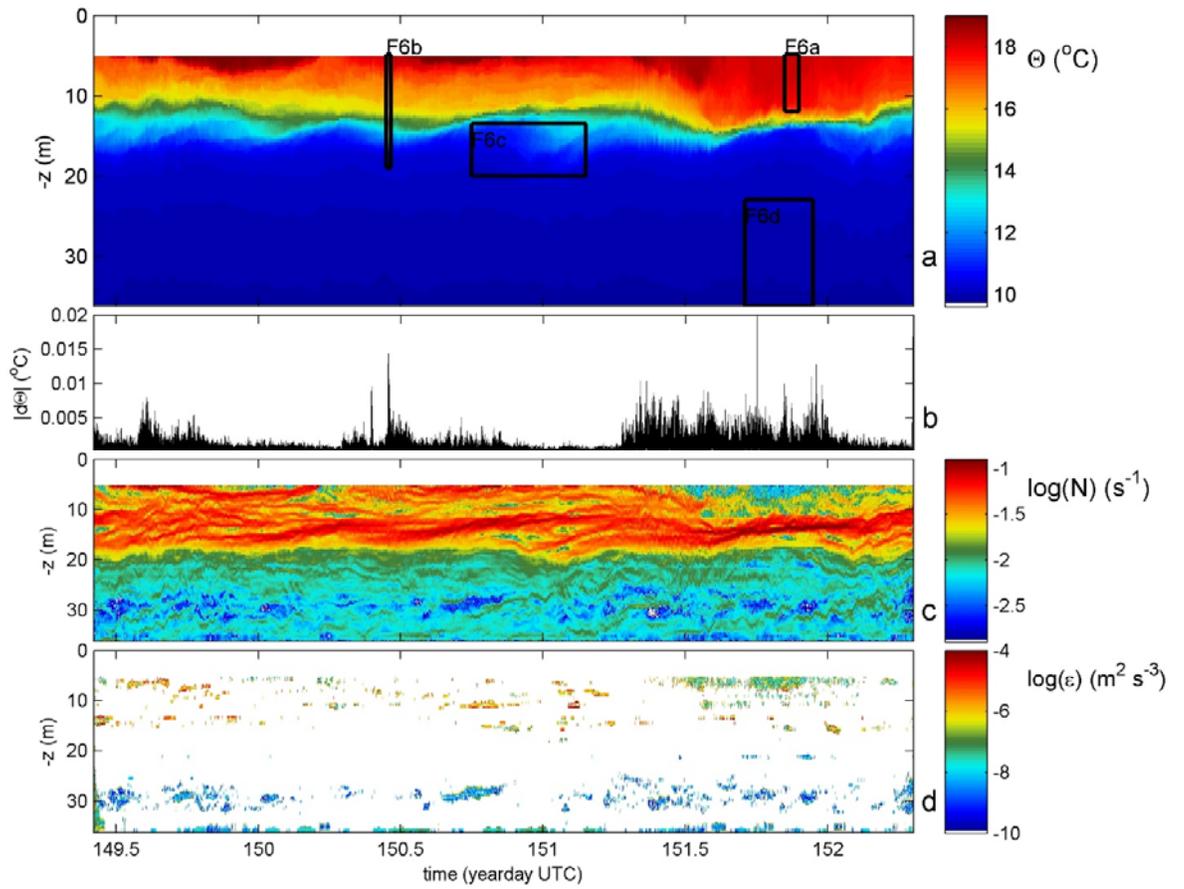

**Fig. 3**. Moored T-sensor data overview of three-day time or time-depth series. (a) Conservative Temperature. The horizontal axis is at the local lake floor. Rectangles refer to detailed plots in Fig. 6a-d. (b) Absolute first derivative of vertical mean values of the data in a. (c) Logarithm of the buoyancy frequency determined from the data in a. after reordering to statically stable profiles. (d) Logarithm of turbulence dissipation rate computed from the data in a.

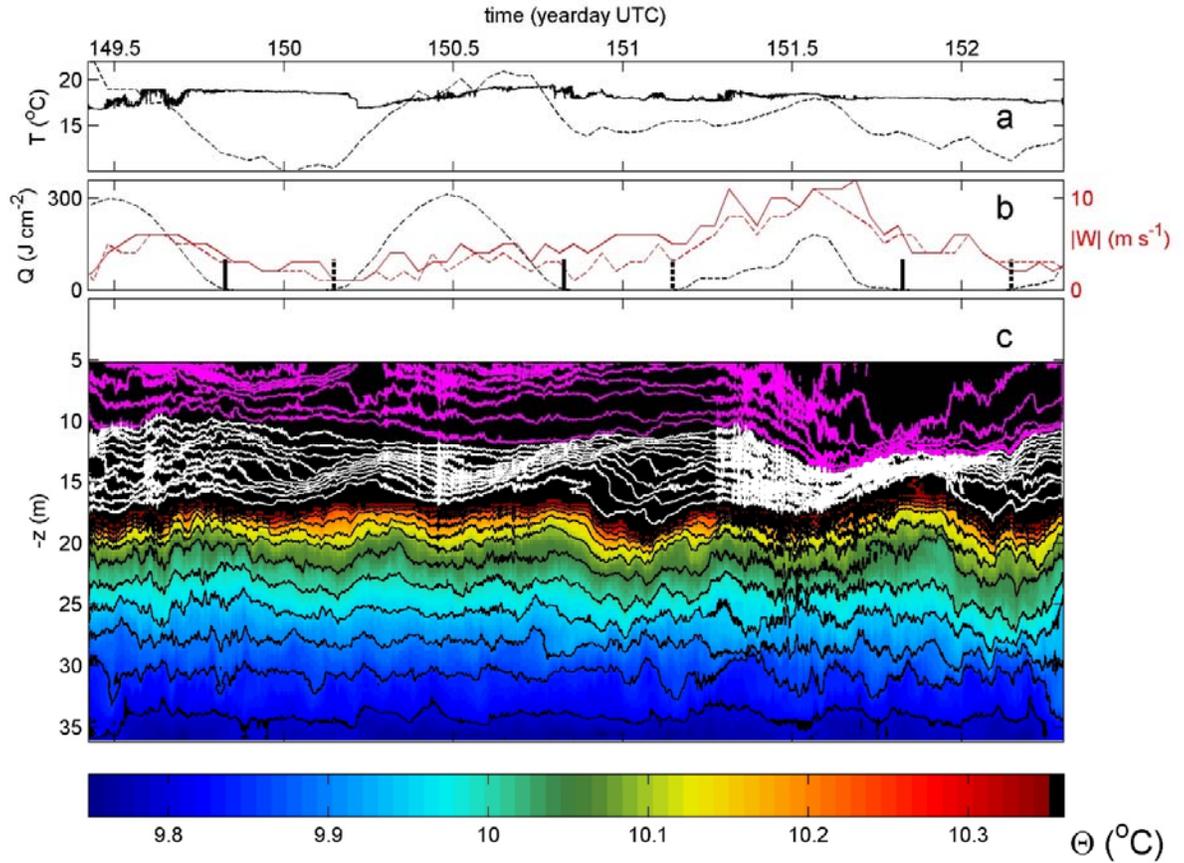

**Fig. 4**. Comparison of nearby weather station information with moored T-sensor data. (a) Temperature measured in water by the upper T-sensor at 5 m below the surface (solid graph) and in air at Wilhelminadorp 'WD' station (dashed). (b) Solar global radiation at WD (dashed black, left-scale) and wind speed (red, right-scale) at WD (dashed) and Vlissingen station (solid). The vertical bars indicate times of local sunrise (solid) and sunset (dashed). (c) As Fig. 3a, but with color-scale focusing on the lower half of the water column and including contours of constant temperature. Solid contours are drawn every 0.05°C, white (distinguishing the pycnocline) and purple (upper layer) contours every 0.5°C.

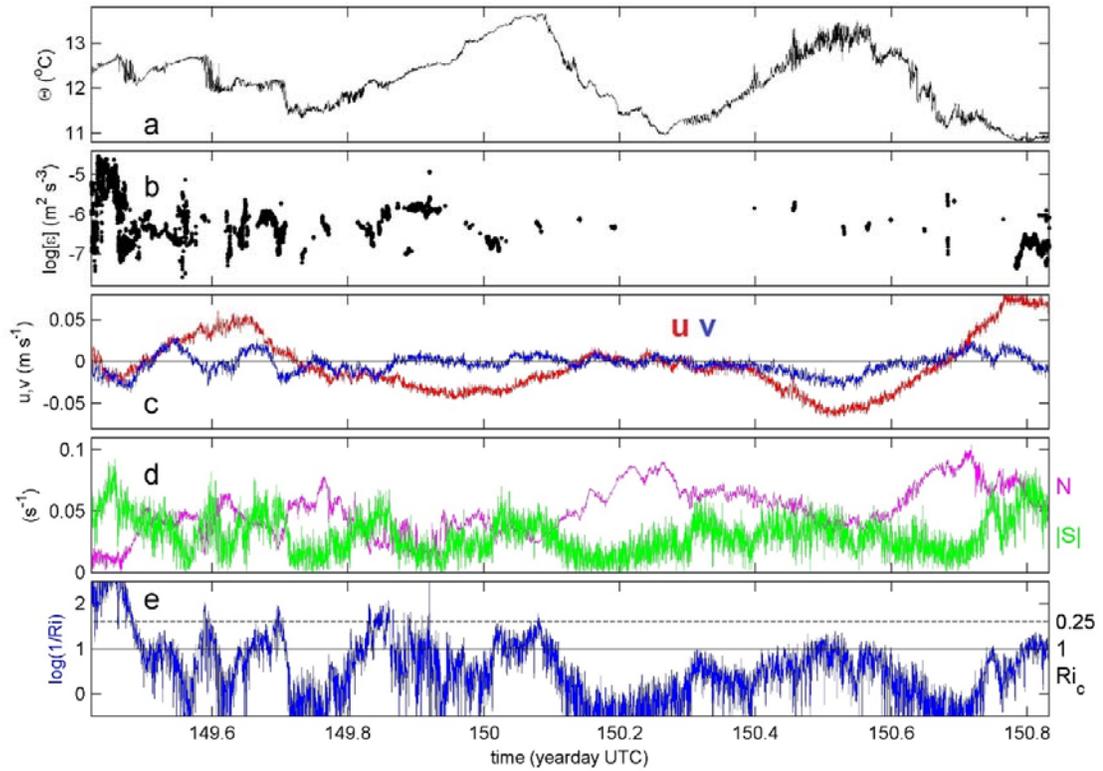

**Fig. 5.** Time series of the first 1.4 days of data when the current profiler at z = -15 m was working. Current meter data are smoothed to 60-s-data for noise removal using a double elliptic filter. (a) Conservative Temperature from T-sensor at z = -14.5 m. (b) Logarithm of turbulence dissipation rate vertically averaged over -16.2 < z < -12.9 m. (c) Along-channel east-west horizontal current component in red and cross-channel component in blue at z - 14.5 m. (d) Buoyancy frequency N from vertically reordered T-sensor data and shear-magnitude |**S**| between -14 < z < -13.5 m. (e) Logarithm of 1/Gradient Richardson number from the data in d., see text. Critical values are indicated by horizontal lines. In this plot unstable portions are above the critical lines.

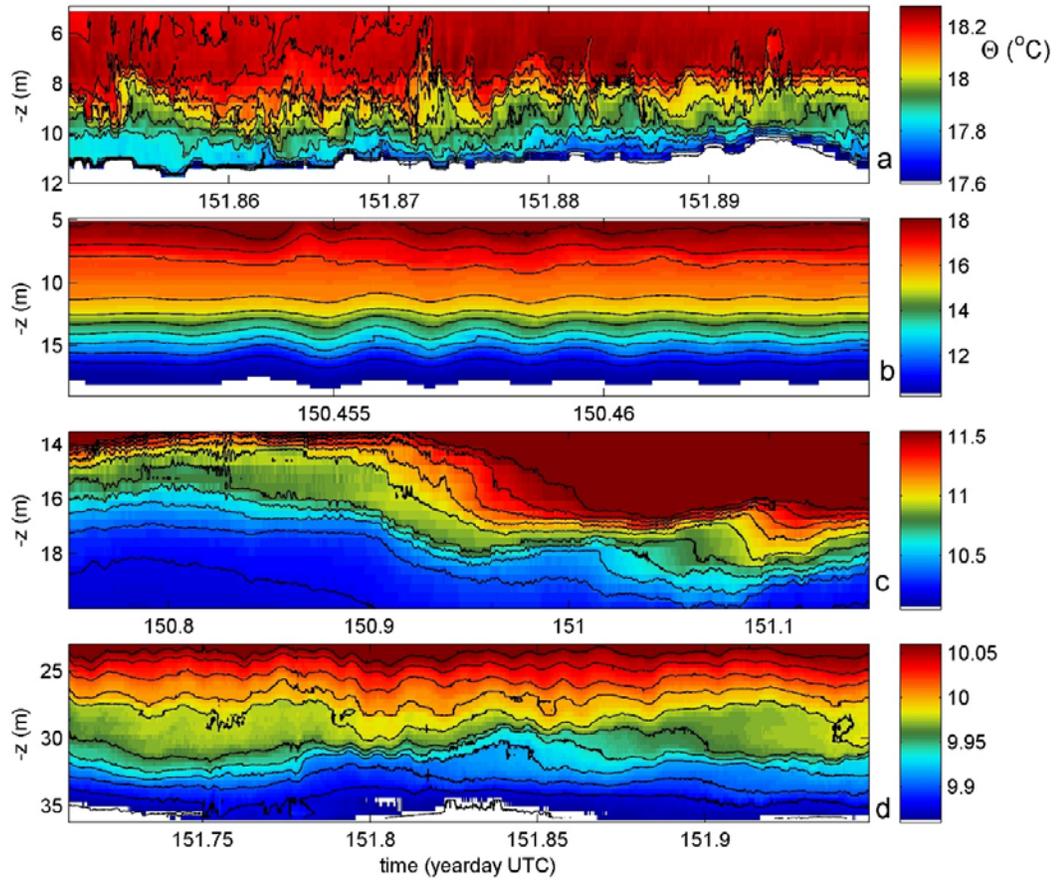

**Fig. 6**. Detail examples of T-sensor data, see Fig. 3 for time-depth ranges. Every panel has different Conservative Temperature (colour) range. Ten black contours are evenly distributed over the temperature range per panel. (a) Near-surface turbulence due to night-time cooling. The time, vertical range of the panel is [1.2 h, 7 m]. (b) High-frequency interfacial wave packet with frequencies near the local buoyancy frequency, [2300 s, 14 m]. (c) Episodic interfacial overturning, [9 h, 6.5 m]. (d) Occasional near-bottom overturning, [6 h, 13 m].